\documentclass[12pt]{article}
\usepackage{amsmath}
\usepackage{graphicx,psfrag,epsf}
\usepackage{enumerate}
\usepackage{natbib}
\usepackage{amsfonts}
\usepackage{multirow}
\usepackage{url} % not crucial - just used below for the URL 
\newcommand{\blind}{1}

\newcommand{\ex}[1]{\ensuremath{\mathbb{E}[#1]}}
\newcommand{\var}[1]{\ensuremath{\mathrm{Var}[#1]}}

\newcommand{\corr}[1]{\ensuremath{\mathrm{Corr}[#1]}}
\newcommand{\bd}[1]{\ensuremath{\mbox{\boldmath $#1$}}}

\begin{document}

\def\spacingset#1{\renewcommand{\baselinestretch}%
{#1}\small\normalsize} \spacingset{1}

%%%%%%%%%%%%%%%%%%%%%%%%%%%%%%%%%%%%%%%%%%%%%%%%%%%%%%%%%%%%%%%%%%%%%%%%%%%%%%

\if1\blind
{
  \title{\bf A spatio-temporal process-convolution model for quantifying  health inequalities in respiratory prescription rates in Scotland}
  \author{Duncan Lee \thanks{The data were provided by the Information Services Division of NHS Scotland, and the work was funded by the UK Medical Research Council (MRC) grant number MR/L022184/1.}\hspace{.2cm}\\
School of Mathematics and Statistics, University of Glasgow\\Duncan.Lee@glasgow.ac.uk
}
\maketitle
} \fi

\if0\blind
{
  \bigskip
  \bigskip
  \bigskip
  \begin{center}
    {\LARGE\bf A spatio-temporal process-convolution model for quantifying  health inequalities in respiratory prescription rates in Scotland}
\end{center}
  \medskip
} \fi

\bigskip
\begin{abstract}
The rates of respiratory prescriptions vary by GP surgery across Scotland, suggesting there are sizeable health inequalities in respiratory ill health across the country. The aim of this paper is to estimate the magnitude, spatial pattern and drivers of this spatial variation. Monthly data on respiratory prescriptions are available at the GP surgery level, which creates an interesting methodological challenge as these data are not the classical geostatistical, areal unit or point process data types. A novel process-convolution model is proposed, which  extends existing methods by being an adaptive smoother via a random weighting scheme and using a tapering function to reduce the computational burden. The results show that particulate air pollution, poverty and ethnicity all drive the health inequalities, while there are additional regional inequalities in rates after covariate adjustment.
\end{abstract}

\noindent%

\vfill

\newpage
\spacingset{1.45} % DON'T change the spacing!

\section{Introduction}
Respiratory disease is a major problem in the United Kingdom, with an estimated  5.4 million people receiving treatment for asthma (\url{https://www.asthma.org.uk}), while  1.2 million have been diagnosed with chronic obstructive pulmonary disease (COPD, \url{https://statistics.blf.org.uk/copd}). Asthma and COPD have no known cure, but the symptoms of breathlessness, coughing and wheezing can usually be managed via a combination of inhalers that either prevent the symptoms from  occurring (using corticosteroids) or relieve symptoms when they occur (using $\beta2$-agonists). These inhalers are obtained by patients via a prescription, which is given to them following a consultation with their general practitioner (GP, doctor). Respiratory medicines (mainly to treat asthma and COPD) account for 13\% of all  prescription costs in Scotland (\citealp{respscotland2014}), making both diseases a major financial and a public health burden.

However, this health burden is not felt uniformly across the country, as there are sizeable spatial inequalities in prescription rates which help  perpetuate an uneven society and lead to vast regional differences in spending by the National Health Service (NHS) Scotland. Quantifying these regional inequalities is important because health care spending is managed in Scotland by 14 regional health boards, which are allocated funding from NHS Scotland via the Resource Allocation Formula (\url{http://www.isdscotland.org/Health-Topics/Finance/Resource-Allocation-Formula/}). Therefore the aim of this paper is to develop a statistical model for explaining the spatio-temporal variation in respiratory prescription rates, particularly focusing on: (i) what are the drivers of the spatial variation; and (ii) how big are the regional variations across the 14 health boards? Previous studies have identified a number of possible drivers of variation in respiratory disease rates, such as particulate air pollution, poverty and ethnicity (see \citealp{partridge2000, sofianopoulou2013,blangiardo2016}), and here we investigate the extent to which these factors impact the spatial variation in respiratory prescription rates in Scotland. Of particular interest is air pollution, as there are large numbers of studies that have linked it to more severe respiratory conditions such as hospital admissions and deaths (e.g. \citealp{lee2008} and \citealp{rushworth2015}). The main question here is does it have similar effects on less severe respiratory disease endpoints such as prescription rates?

Modelling these data provides a statistical challenge, as GPs who write the prescriptions are grouped into 966 GP surgeries that each have a single geographical coordinate. Therefore these data are not strictly of a  geostatistical, areal unit, or point process data type, which is the commonly used typology of spatial data. Geostatistical data models assume that the process being modelled is in theory measurable everywhere, areal unit data models require the data to relate to non-overlapping areal units, while point process models assume the locations of the data points are random. None of these assumptions are realistic here, as the GP surgery locations are fixed, do not exist at unmeasured locations, and do not sub-divide the population into spatially non-overlapping groups. Therefore the use of standard Gaussian random field (GRF, \citealp{diggle2007}), Gaussian Markov random field (GMRF, \citealp{rue2005}) or log-Gaussian Cox process (LGCP, \citealp{diggle2013}) models commonly used for these 3 spatial data types is inappropriate. Existing spatial analyses of GP prescription data have been undertaken recently by  \cite{rowlinson2013, sofianopoulou2013} and \cite{blangiardo2016}, but while the first applies a kernel smoothing algorithm to the data in a non-regression context, the latter two aggregate the GP practices to non-overlapping  areal units and respectively utilise mixed effects and  GMRF models to analyse the resulting areal unit data.

Therefore this paper proposes a novel process-convolution based approach for modelling this non-standard spatial data type, which does not make the unrealistic assumptions outlined above about the data locations. The model builds on the original proposal by \cite{higdon1998} for univariate geostatistical climate  data, but we note that the model is very similar in form to both spatial moving average models (see \citealp{best2005}), and kernel smoothers (see \citealp{bowman1997}). These models  represent the data as a weighted average of white noise, where the weights are defined by a Kernel function depending on the distance between the white noise process and the data,  modulated by a bandwidth parameter. However, the exploratory graphical analysis in Section 2 shows that using a common distance-decay Kernel function with a single bandwidth parameter is likely to be inappropriate for the GP prescription data, because in addition to exhibiting spatial smoothness, there are many boundaries (abrupt step changes) in prescription rates between spatially close GP surgeries. In his original paper \cite{higdon1998} allowed for varying levels of spatial smoothness by using  different bandwidth parameters in different spatial subregions, but here we want to allow for some pairs of GP surgeries to be spatially close but have very different values corresponding to little or no correlation between them. rather than allowing for different levels of spatial smoothness in the study region.

Therefore we propose a novel Bayesian random weighting spatio-temporal process-convolution model,  with inference based on Markov chain Monte Carlo (MCMC) simulation. The model is an adaptive spatial smoother, in that pairs of geographically close data points can be modelled as autocorrelated or marginally independent, allowing boundaries between spatially close data points to be identified. Adaptive smoothing models allowing boundaries to be identified in areal unit data have received much attention in the disease mapping literature, with examples including \cite{lu2007, lee2012} and \cite{rushworth2016}. However, their development within the class of process-convolution models is under researched and is the methodological contribution of this paper. Instead advances in process-convolution modelling have focused on  intrinsic rather than white noise processes (\citealp{lee2005}), temporally autocorrelated dynamic factors (\citealp{calder2007}),  and interpolation between spatial frameworks (\citealp{congdon2014}).

Our proposed methodology relaxes the assumption of fixed convolution weights based on a Kernel function, and instead assigns the weights Dirichlet priors and  estimates them in the model. However, process-convolution models can be computationally expensive to implement, so we propose a novel computationally efficient tapering approach that makes the matrix of weights sparse, which is similar in spirit to covariance tapering (\citealp{kaufman2008}).  The model and its implementation via freely available software is presented in Section 3, following the data description and exploratory analysis in Section 2. Section 4 presents the results of analysing the GP prescription data, and compares the random weighting model proposed here against a simpler distance-decay alternative. Finally, the paper concludes with a discussion in Section 5.

\section{Data and exploratory analysis}

\subsection{Study region}
The study region is Scotland, and a map showing the main settlements is displayed in Figure 1 of the supplementary material accompanying this paper. Scotland has a population of around 5.4 million people, who are served by 966 GP surgeries. These surgeries are arranged within 14 regional health boards, through which health care spending is managed. The locations of the surgeries and of the health board is displayed in Figure 2 in the supplementary material, and here we only model data from 948 of the 966 surgeries due to missing data problems (see below). The number of surgeries per health board is shown in Table 1 in the supplementary material, and the largest numbers of surgeries are in Greater Glasgow and Clyde (241 surgeries) and Lothian (120 surgeries), which contain the two largest cities Glasgow and Edinburgh and have populations of around 1.2 million and 0.85 million respectively. These health boards also have a high density  of GP surgeries, which contrasts with the mostly rural highlands health board that has large un- or sparsely populated areas with no GP surgeries.\\

\subsection{Prescription data}
Newly relesased data on respiratory prescriptions  were obtained from the Information Services Division of NHS Scotland (\url{http://www.isdscotland.org/Health-Topics/Prescribing-and} \url{-Medicines/Publications/2016-11-15/opendata.asp}), and are available at a monthly resolution from October 2015 to June 2016 inclusive. We consider respiratory prescriptions for inhalers that relieve the symptoms of asthma and COPD, namely those containing  short acting $\beta_2$ agonists. Specifically, for each GP surgery and month we have the total number of prescriptions for  Salbutamol (100$mcg$ or 200$mcg$) and Ventolin (100$mcg$ or 200$mcg$), which are the two most commonly used respiratory medications.

The number of prescriptions at each GP surgery will depend on the size and age-sex structure of its patient population (list size). This is accounted for by computing the expected numbers of respiratory disease sufferers (from asthma or copd) registered with each GP surgery using  indirect standardisation. Specifically, we have data on the numbers of males and females at each surgery in the following age groups: 0-4, 5-14, 15-24, 25-44, 45-64, 65-74, 75-84, 85+. These surgery age-sex list sizes are multiplied by age-sex specific Scotland-wide rates of asthma and COPD, before being summed over the groups to obtain an expected number of patients with respiratory disease at each surgery. These expected numbers are  then scaled so that over all surgeries and months the total observed numbers  of prescriptions equals the total expected numbers. However,  18 GP surgeries had missing age-sex specific list size data, and thus they were removed from the data set.

\begin{figure}
\centering 
\scalebox{0.3}{\includegraphics{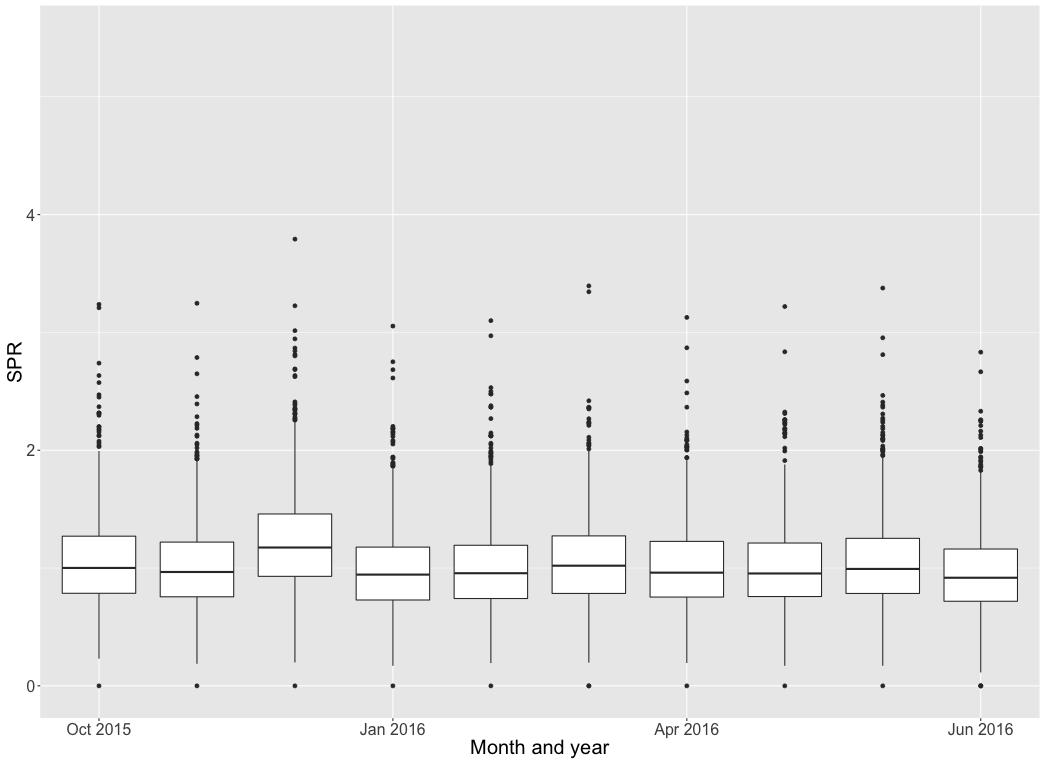}}
  \caption{Boxplots showing the monthly variation in the SPR distribution.
  \label{figure2}}
\end{figure}

An exploratory measure of the prescription rate for each surgery and month is the observed number of prescriptions divided by the scaled expected number of patients with respiratory disease, which is similar to a standardised mortality ratio (SMR) when modelling death rates. We term this ratio the standardised prescription rate (SPR), and a rate of 1.2 means a surgery has a 20\% increased rate of prescriptions compared to the Scotland average. The distributions of the raw prescription counts and the SPR are summarised in Table \ref{table1}, which shows that there are some extreme high values as well as some GP surgeries with zero counts for some months. The monthly temporal pattern in SPR is displayed by boxplots in Figure \ref{figure2}, and shows little change in the distributions over time except for December which has a noticeably higher rate than the other months.

\begin{figure}
\centering\caption{Maps showing the spatial distribution in average SPR over all 10 months for (a) Scotland and (b) the central belt containing Glasgow in the west and Edinburgh in the east. To aid visualisation the SPR has been categorised as: Low - $<0.9$, Average - $[0.9, 1.1]$, and High - $>1.1$.}
\label{figure3}
\begin{picture}(10,16)
\put(1,4.5){\scalebox{0.30}{\includegraphics{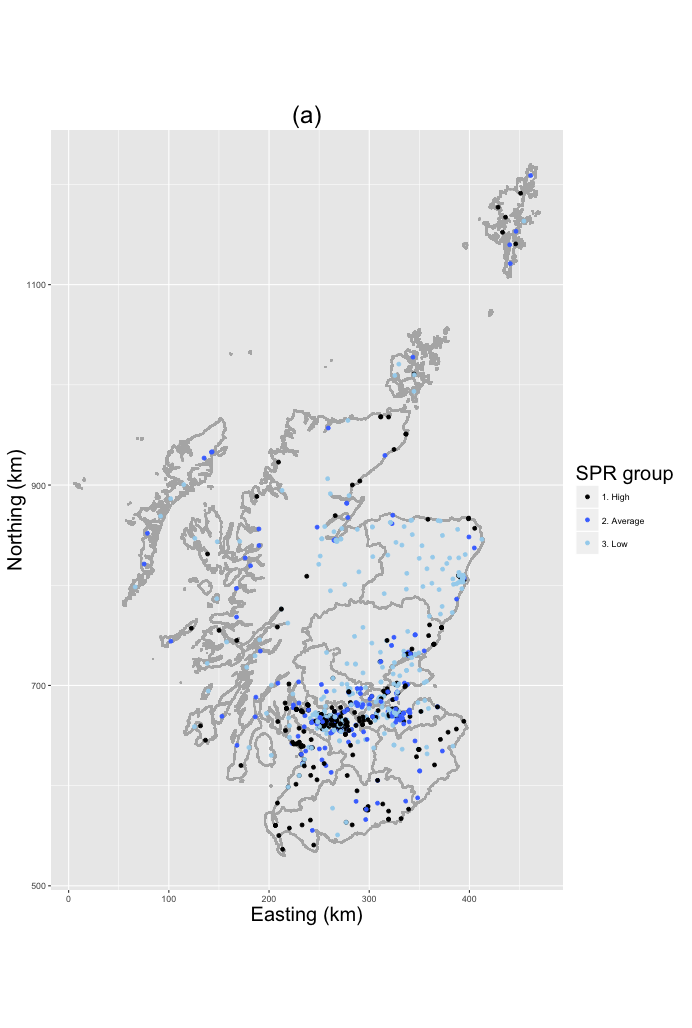}}}
\put(-1,0){\scalebox{0.35}{\includegraphics{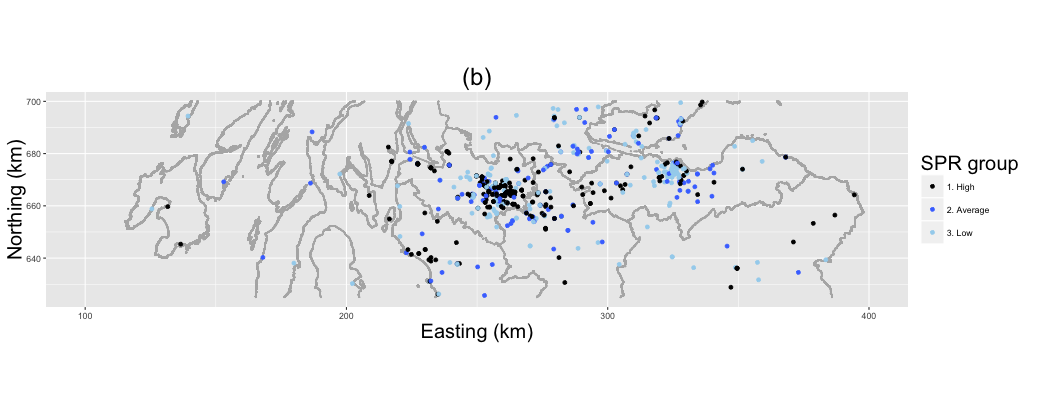}}}
\end{picture}
\end{figure}

The average spatial pattern in the SPR over all 10 months is displayed in Figure \ref{figure3}, where to allow the spatial pattern to be clearly seen the SPR has been classified into 3 groups: Low - $<0.9$, Average - $[0.9, 1.1]$, and High - $>1.1$. A continuous SPR colour scale was originally used, but the resulting  colour variation was hard to see, hence the use of  a discrete scale. Panel (a) relates to the whole of Scotland, while panel (b) zooms in on the central belt containing Glasgow in the west and Edinburgh in the east, the 2 largest cities (see Figure 1 in the supplementary material). The figure shows the main areas with high rates are the health boards of Greater Glasgow and Clyde, and Shetland, where as low rates are observed in Grampian and Lothian. The figure also shows spatial autocorrelation in the SPR, with large groups of the same coloured points close together. However, the figure also shows that there are numerous geographically close pairs of surgeries with respectively high and low prescription rates, suggesting that the spatial autocorrelation structure is not globally smooth and instead contains boundaries that correspond to abrupt step changes. This observation motivates our random weighting methodology outlined in the next section, and in Section 4 we show that it fits the data better than simple distance-decay weights.

\subsection{Covariates}
Covariate information was collected to understand the drivers of the spatial variation in the GP prescription rates. The available covariates varied spatially by GP surgery but were static in time, because no monthly varying covariate information could be obtained. GP surgery specific poverty and ethnicity information is largely unavailable, so instead data were obtained on the assumption that the surgery patient population is well represented by the population living in the closest small area administrative unit for which data are available. Poverty is measured by the median property price in the closest intermediate zone, while the proportion of patients that are white is estimated from the census (in 2011) in a similar manner. Finally, air pollution is well known to impact respiratory disease, so here we obtain modelled concentrations of particulate matter less than 10 (PM$_{10}$) and 2.5 (PM$_{2.5}$) microns in size measured in $\mu gm^{-3}$ for 2014 (the closest year available). These concentrations relate to 1km grid squares, and can be obtained from \url{https://uk-air.defra.gov.uk/data/pcm-data} (\cite{steadman2015}) for the entire United Kingdom. Here, the grid square containing the GP surgery was used as the exposure.

\begin{table}
\centering\begin{tabular}{llrrrr}
\hline
\raisebox{-1.5ex}[0pt]{\textbf{Variable}} &\multicolumn{5}{c}{\textbf{Percentile of the distribution}}\\
&\textbf{0\%}&\textbf{25\%}&\textbf{50\%} &\textbf{75\%}&\textbf{100\%}\\\hline
Prescriptions (totals) & 0   &84  &140  &210  &772 \\
SPR &0.00 &0.77 &0.99 &1.25 &3.79\\\hline
PM$_{2.5}$ ($\mu gm^{-3}$)&4.36  &7.06 & 7.87  &8.49 &16.52 \\
PM$_{10}$ ($\mu gm^{-3}$) &6.17  &9.87 &11.06 &12.07 &23.39\\
Property price ($\pounds 000$) &34.00  &77.00 &105.81 &147.50 &375.01\\
Proportion white &0.36 &0.94 &0.98 &0.99 &1.00\\\hline
\end{tabular}
\caption{Summary of the distribution of the non-categorical variables in the study.} \label{table1}
\end{table}

\section{Methodology}
This section describes the adaptive spatio-temporal process-convolution methodology  proposed in this paper, as well as outlining a simplified globally smooth distance-decay alternative. The model is described in Sections 3.1 and 3.2, its theoretical properties are outlined in Section 3.3, while software to fit the model is discussed in Section 3.4.

\subsection{Overall data model}
Let $\mathbf{s}_k=(s_{1k}, s_{2k})$ denote the geographical coordinate (easting, northing) of the $k$th GP surgery, where $k=1,\ldots,K=948$. Then let $(Y_{t}(\mathbf{s}_k), E_{t}(\mathbf{s}_k))$ respectively denote the observed number of prescriptions and the scaled expected number of respiratory disease patients for the $k$th GP surgery  and $t$th month ($t=1,\ldots,N=10$). Finally, let $\mathbf{x}_t(\mathbf{s}_k)=(x_{t1}(\mathbf{s}_k),\ldots,x_{tp}(\mathbf{s}_k))$ denote a $p\times 1$ vector of covariates for the $k$th surgery in the $t$th month including a column of ones for the intercept term. Then the first level of the model proposed here is given by

\begin{eqnarray}
Y_t(\mathbf{s}_k)&\sim&\mbox{Poisson}(E_{t}(\mathbf{s}_k)R_{t}(\mathbf{s}_k))~~~~\mbox{for }k=1,\ldots,K,~t=1,\ldots,N,\label{likelihood}\\
\ln(R_t(\mathbf{s}_k))&=& \mathbf{x}_t(\mathbf{s}_k)^{\top}\bd{\beta} + \sum_{j=1}^{K}w_{kj}\theta_t(\mathbf{s}_j),\nonumber\\
\bd{\beta}&\sim&\mbox{N}(\bd{\mu}_{\beta}, \bd{\Sigma}_{\beta}).\nonumber
\end{eqnarray}

Here $R_{t}(\mathbf{s}_k)$ denotes the rate of prescriptions compared to the scaled expected numbers, and has the same interpretation as the simple  SPR statistic outlined above. The log rate is modelled by covariates and a process-convolution latent process, where the regression parameters for the former are assigned a Gaussian prior with weakly informative hyperparameters ($\bd{\mu}_{\beta}=\mathbf{0}$ - a vector of zeros, $\bd{\Sigma}_{\beta}=1000\mathbf{I}_p$ - where $\mathbf{I}_p$ is the $p\times p$ identity matrix). The latent variable $\phi_{t}(\mathbf{s}_k)=\sum_{j=1}^{K}w_{kj}\theta_t(\mathbf{s}_j)$ is included to model any unmeasured spatio-temopral autocorrelation in the data after covariate adjustment, which would likely be induced by factors such as unmeasured confounding.

We take a process-convolution approach here, because as discussed in the introduction geostatistical GRF, areal unit GMRF, and point process LGCP models make inappropriate assumptions about the data locations. For month $t$ the process-convolution $\phi_{t}(\mathbf{s}_k)$  is based on  $K$ spatially unstructured random variables $(\theta_t(\mathbf{s}_1),\ldots,\theta_t(\mathbf{s}_K))$ centred at the data locations, which is similar to the approach taken in Kernel smoothing (\citealp{bowman1997}). Spatial autocorrelation is induced into the model by the $K\times K$ matrix of weights $\mathbf{W}=(w_{kj})_{K\times K}$, and further details are provided in Section 3.2 below. In contrast, temporal autocorrelation is induced  by assigning each $\theta_t(\mathbf{s}_j)$ an independent first order autoregressive process prior:

\begin{eqnarray}
\theta_t(\mathbf{s}_j)&\sim&\mbox{N}(\gamma\theta_{t-1}(\mathbf{s}_j),\tau^2)~~~~\mbox{for }t=2,\ldots,N,~j=1,\ldots,K,\label{ar1}\\
\theta_1(\mathbf{s}_j)&\sim & \mbox{N}(0,\tau^2),\nonumber\\
\gamma &\sim & \mbox{Uniform}(0,1)\nonumber,\\
\tau^{2}&\sim&\mbox{Inverse-Gamma}(a=1,b=0.01).\nonumber
\end{eqnarray}

Here $\gamma$ controls the temporal autocorrelation in the spatially unstructured latent process, with $\gamma=0$ corresponding to temporal independence, while $\gamma=1$ corresponds to a non-stationary temporally dependent random walk model.

\subsection{Modelling the weights $\mathbf{W}$}
The simplest approach to modelling   $\mathbf{W}$ is to use Kernel-based weights, which if a Gaussian kernel function is used have the form:

\begin{equation}
w_{kj}=\frac{\frac{1}{\sqrt{2\pi/\alpha}}\exp\left(-\alpha\frac{||\mathbf{s}_k -\mathbf{s}_j||}{2}\right)}{\sum_{i=1}^{K}\frac{1}{\sqrt{2\pi/\alpha}}\exp\left(-\alpha\frac{||\mathbf{s}_k -\mathbf{s}_i||}{2}\right)},\label{weight0}
\end{equation}

where $||.||$ denotes Euclidean distance. Thus the closer two $(\mathbf{s}_k,\mathbf{s}_j)$ locations are the larger the weight. The speed at which the weights decay to zero with increasing distance is controlled by $\alpha$, which is known as the bandwidth parameter. If $\alpha$ is very small the weights are non-negligible even for relatively large distances, resulting in a spatially smooth surface with a large range (data points far away will be autocorrelated). Conversely, as $\alpha$ increases only nearby $(\mathbf{s}_k,\mathbf{s}_j)$ locations have non-negligible weights, resulting in only short range autocorrelation. In the limit as $\alpha\rightarrow\infty$ then $w_{kk}\rightarrow 1$ for all $k$ and $w_{kj}\rightarrow 0$ for all $k\neq j$, resulting in spatial independence between all pairs of data points. However, this specification results in $\mathbf{W}$ being a dense matrix, which makes the MCMC estimation algorithm computationally demanding. To see this consider the full conditional distribution of $\theta_t(\mathbf{s}_j)$:

\footnotesize\begin{eqnarray}
f(\theta_t(\mathbf{s}_j)|-)&\propto&\left[\prod_{k=1}^{K}\exp\left(-E_{t}(\mathbf{s}_k)\exp(\mathbf{x}_t(\mathbf{s}_k)^{\top}\bd{\beta} + \sum_{j=1}^{K}w_{kj}\theta_t(\mathbf{s}_j))\right)\left(E_{t}(\mathbf{s}_k)\exp(\mathbf{x}_t(\mathbf{s}_k)^{\top}\bd{\beta} + \sum_{j=1}^{K}w_{kj}\theta_t(\mathbf{s}_j))\right)^{Y_t(\mathbf{s}_k)}\right]\nonumber\\
&\times &\exp\left(-\frac{1}{2}\left[\frac{(\theta_t(\mathbf{s}_j)-\gamma \theta_{t-1}(\mathbf{s}_j))^2+(\theta_{t+1}(\mathbf{s}_j)-\gamma \theta_{t}(\mathbf{s}_j))^2}{\tau^2}\right] \right),\label{fc}
\end{eqnarray}
\normalsize

where `$-$' denotes the data and all other relevant parameters. Thus when updating $\theta_t(\mathbf{s}_j)$ one must evaluate the data likelihood for all $K$ data points at time $t$, which when iterated over all $K\times N$ elements $\{\theta_t(\mathbf{s}_j)\}$ for a large number of MCMC iterations is computationally demanding. To reduce this demand we propose a novel tapering approach that makes the weight matrix $\mathbf{W}$ sparse, using an idea similar to covariance tapering (\citealp{kaufman2008}). Consider the tapering function:

\begin{equation}
I(\mathbf{s}_k, \mathbf{s}_j)=\left\{\begin{array}{ll}1&\mbox{if location $\mathbf{s}_j$ is one of the $m$ closest points to location $\mathbf{s}_k$}\\0&\mbox{Otherwise}\\
\end{array}\right.,\nonumber
\end{equation} 

which can be combined with (\ref{weight0}) to create the following sparse weights $\mathbf{W}=(w_{kj})$:

\begin{equation}
w_{kj}=\frac{I(\mathbf{s}_k, \mathbf{s}_j)\frac{1}{\sqrt{2\pi/\alpha}}\exp\left(-\alpha\frac{||\mathbf{s}_k -\mathbf{s}_j||}{2}\right)}{\sum_{i=1}^{K}I(\mathbf{s}_k, \mathbf{s}_i)\frac{1}{\sqrt{2\pi/\alpha}}\exp\left(-\alpha\frac{||\mathbf{s}_k -\mathbf{s}_i||}{2}\right)}.\label{weight1}
\end{equation}

These weights follow the Kernel density form except that they have been tapered, so that only the  $m$ spatially closest elements $\mathbf{s}_{j}$ to $\mathbf{s}_k$ have a non-zero weight  $w_{kj}$. The resulting $\mathbf{W}$ matrix is sparse with only $mK$ non-zero elements, which means that when updating $\theta_t(\mathbf{s}_j)$ using (\ref{fc})  the data likelihood only needs to be evaluated at a small number of data points $Y_{t}(\mathbf{s}_k)$ rather than at all $K$. A uniform prior can be specified for $\alpha$, and the model can be fitted with different amounts of tapering (different values of $m$) to see its impact on model performance (as is done in Section 4). 

The weights given by (\ref{weight1}) are globally smooth, in the sense that a single bandwidth parameter $\alpha$ controls the entire $\mathbf{W}$ matrix. Thus if $\alpha$ is small then the spatial surface will have long-range autocorrelation everywhere, where as if $\alpha$ is large it will only exhibit short-range autocorrelation everywhere. However, as seen in Section 2 some pairs of geographically close GP surgeries have similar prescription rates suggesting spatial autocorrelation, while other geographically close pairs have very different values suggesting independence. This localised autocorrelation structure cannot be well captured by (\ref{weight1}), as if two pairs of points are the same distance apart their weights should be similar regardless of the similarity of their data values. Note, they will not be exactly the same due to the uneven spatial configuration of the GP surgeries. To remedy this we propose a novel random weighting scheme for $\mathbf{W}$, which when combined with the tapering function is given by:

\begin{eqnarray}
w_{kj}&=&\left\{\begin{array}{ll}\psi_{kr}&\mbox{if $\mathbf{s}_j$ is the $r$th closest point to $\mathbf{s}_k$ for $r=1,\ldots,m$}\\0&\mbox{Otherwise}\\
\end{array}\right., \label{weight2}\\
\bd{\psi}_{k}=(\psi_{k1},\ldots, \psi_{km})&\sim &\mbox{Dirichlet}(\alpha_1=1,\ldots,\alpha_m=1).\nonumber
\end{eqnarray}

Spatial autocorrelation is retained via the taper function, because only the $m$ geographically closest $\theta_{t}(\mathbf{s}_j)$ values are given non-zero weight in the process-convolution $\phi_{t}(\mathbf{s}_k)$. However, the use of random weights means that spatially close elements $(\phi_{t}(\mathbf{s}_k), \phi_{t}(\mathbf{s}_i))$ can be highly autocorrelated if the weights $(w_{ki}, w_{ik})$ are high, but can also be close to independent if close to zero weights are estimated. This model also encompasses global smoothness and independence as special cases, as weights from  (\ref{weight1}) are clearly special cases of this  general framework. A Dirichlet prior is assumed for each set of weights $\bd{\psi}_{k}$ so that they sum to one, and we note that to aid parameter identifiability in the MCMC estimation algorithm the same set of weights are used for all time periods ($\bd{\psi}_{k}$ does not depend on $t$). This simplification gives temporal replication in the spatial surface and hence more information with which to estimate the weights, as the number of weights $K\times m$ is larger than the number of spatial data points $K$.

\subsection{Autocorrelation structure assumed by the model}
The autocorrelation structure between two random effects $(\phi_{t}(\mathbf{s}_k), \phi_{r}(\mathbf{s}_i))$ can be obtained using multivariate Gaussian theory as follows. First, the  autoregressive process for $\bd{\theta}(\mathbf{s}_j)=(\theta_1(\mathbf{s}_j),\ldots,\theta_N(\mathbf{s}_j))$ given by (\ref{ar1}) can be re-written as $\bd{\theta}(\mathbf{s}_j)\sim\mbox{N}(\mathbf{0}, \tau^2\mathbf{Q}^{-1})$, where $\mathbf{0}$ is an $N\times 1$ vector of zeros and the precision matrix $\mathbf{Q}_{N\times N}$ is tridiagonal and given by:

\begin{eqnarray}
\mathbf{Q}~=~\left(\begin{array}{lllll}
1+\gamma^2 & -\gamma &  & & \\
-\gamma & 1+\gamma^2 & -\gamma & & \\
&\ddots & \ddots & \ddots & \\
&&-\gamma &1+\gamma^2 &-\gamma\\
&&&-\gamma&1\\
\end{array}\right).\label{ar_precision}
\end{eqnarray}

Then letting $\bd{\theta}=(\bd{\theta}(\mathbf{s}_1),\ldots,\bd{\theta}(\mathbf{s}_K))$ it follows that

\begin{equation}
\bd{\theta}~\sim~\mbox{N}\left(\mathbf{0}, \tau^2[\mathbf{I}_K \otimes\mathbf{Q}]^{-1}\right),\label{thetadist}
\end{equation}

where $\mathbf{I}_K$ is the $K\times K$ identity matrix and $\otimes$ is the Kronecker product. Then writing $\bd{\phi}=(\bd{\phi}(\mathbf{s}_1),\ldots,\bd{\phi}(\mathbf{s}_K))$, where $\bd{\phi}(\mathbf{s}_j)=(\phi_1(\mathbf{s}_j),\ldots,\phi_N(\mathbf{s}_j))$, it follows that $\bd{\phi}=(\mathbf{W}\otimes\mathbf{I}_N)\bd{\theta}$, where $\mathbf{I}_N$ is the $N\times N$ identity matrix. Then using (\ref{thetadist}) we have that:

\begin{equation}
\bd{\phi}~\sim~\mbox{N}\left(\mathbf{0}, \tau^2[\mathbf{W}\otimes\mathbf{I}_N][\mathbf{I}_K \otimes\mathbf{Q}]^{-1}[\mathbf{W}\otimes\mathbf{I}_N]^{\top}\right)~=~\mbox{N}\left(\mathbf{0}, \tau^2[\mathbf{W}\mathbf{W}^{\top}\otimes\mathbf{Q}^{-1}]\right).\label{phidist}
\end{equation}

Thus $\ex{\phi_{t}(\mathbf{s}_k)}=0$ for all $(k,t)$, whilst the variance, temporal and spatial autocorrelations simplify to:

\begin{eqnarray}
\var{\phi_{t}(\mathbf{s}_k)}&=&\tau^2[\mathbf{Q}^{-1}]_{tt}\sum_{j=1}^{K}w_{kj}^2\label{corrandvar},\\
\corr{\phi_{t}(\mathbf{s}_k),\phi_{r}(\mathbf{s}_k)} &=& \frac{[\mathbf{Q}^{-1}]_{tr}}{\sqrt{[\mathbf{Q}^{-1}]_{tt}[\mathbf{Q}^{-1}]_{rr}}},\nonumber\\
\corr{\phi_{t}(\mathbf{s}_k),\phi_{t}(\mathbf{s}_i)} &=& \frac{\sum_{j=1}^{K}w_{kj}w_{ij}}{\sqrt{\left(\sum_{j=1}^{K}w_{kj}^2\right)\left(\sum_{j=1}^{K}w_{ij}^2\right)}}.\nonumber
\end{eqnarray}

The variance depends on both the spatial structure in the data via the weights $\mathbf{W}$ and the temporal structure in the data via $\mathbf{Q}$, and in the simple case of no temporal dependence ($\gamma=0$) then $[\mathbf{Q}^{-1}]_{tt}=1$ and only the spatial structure impacts the variance. The spatial and temporal autocorrelation structures separate, in the sense that the temporal autocorrelation depends only on $\mathbf{Q}$ whilst the spatial autocorrelation depends only on $\mathbf{W}$. The tapering applied to $\mathbf{W}$ means that $(\phi_{t}(\mathbf{s}_k),\phi_{t}(\mathbf{s}_i))$ are marginally independent if the two sets of $m$ closest points to ($\mathbf{s}_k, \mathbf{s}_i$) are disjoint. However, for model (\ref{weight2}) close to independence can also be achieved for a pairs of GP surgeries with non-disjoint  tapering sets, without enforcing the same low or zero correlation on other similar pairs of surgeries. Note, this is not possible under (\ref{weight1}), where the weights depend globally on distance and the bandwidth parameter $\alpha$. Note also that explicit algebraic forms for the marginal autocorrelations cannot be obtained for GMRF models, which instead are parameterised in terms of the precision matrix and hence only partial autocorrelations and conditional independences can be directly inferred.

\subsection{Model fitting and software}
Software to fit the two models presented here  (with weights (\ref{weight1}) and (\ref{weight2}))  using MCMC simulation is provided in the supplementary material accompanying this paper, and uses a combination of Gibbs sampling and Metropolis updating steps. The software is written in the \texttt{R} (\citealp{R}) programming language, and is similar in style to the \texttt{CARBayes} (\citealp{lee2013}) and \texttt{CARBayesST} software packages. The software is made computationally efficient by use of \texttt{C++} subroutines written using the \texttt{Rcpp} (\citealp{Eddelbuettel2011}) package, as well as taking advantage of the sparsity of $\mathbf{W}$ via a triplet form representation. Also provided with the software is the data analysed here and a script file to recreate the results from both models  (\ref{weight1}) and (\ref{weight2}), which both makes this research reproducible and allows other to apply the models to their own data. Also provided is a second script file that simulates data and fits both models, which illustrates the correctness of the algorithms.

\section{Results}
Models with  global (\ref{weight1}) and adaptive (\ref{weight2}) weights were fitted to the respiratory prescriptions data described in Section 2, where the tapering parameter $m$ was set equal to $m=4,8,16$. For all 6 models inference was based on 30,000 MCMC samples generated  from 3 parallel Markov chains, that were each burnt in for 100,000 iterations at which point convergence was assessed to have been reached using a combination of trace-plots of sample parameters and the Geweke statistic (\citealp{Geweke}). After the burn-in period a further 100,000 samples were generated, which were then thinned by 10 to reduce the autocorrelation.  The overall fit and computational speed of fitting each model is summarised below, while the following sections describe the covariate effects, health board  inequalities and the localised nature of the spatial dependence in the data.

\subsection{Model fit}
The overall fit of each model to the data is summarised in Table \ref{modelfit} by both the Watanabe-Akaike Information Criteria (WAIC, \citealp{watanabe2010}) and the Log Marginal Predictive Likelihood (LMPL, \citealp{congdon2005}). In all cases all covariates described in the next section are included in the models, except that PM$_{10}$ is excluded as it is highly correlated with PM$_{2.5}$ (correlation of 0.95). Both statistics strongly suggest that model (\ref{weight2}) with adaptive weights exhibits a much better fit to the data than model (\ref{weight1}) with global weights, as the former has much smaller WAIC statistics and much larger LMPL values. The magnitude of these differences are large, as for example when $m=8$ the WAIC is smaller by 19.9\% (a difference of 18,476.0 in absolute value), while the LMPL is larger by 16.2 \% (a difference of 6,609.7 in absolute value). In contrast, changing $m$ only has a relatively small impact on model fit, with larger values of $m$ fitting the data slightly better than smaller values. For example, the WAIC statistics only change by 0.046\% and 0.60\% for models (\ref{weight1}) and (\ref{weight2}) respectively, as $m$ increases from $m=4$ to $m=16$. However, increasing $m$ vastly increases the computational burden, as the time taken to fit the models increases by 2.89  (for model (\ref{weight1})) and 2.63  (for model (\ref{weight2}))  times as $m$ increases from $m=4$ to $m=16$. In contrast, the increase in computational burden from model (\ref{weight1}) to model (\ref{weight2}) is not large, with for example a 17\% increase when $m=4$.

\begin{table}
\centering\begin{tabular}{lllll}
\hline
\multirow{2}{*}{\textbf{m}}  &\multicolumn{2}{c}{\textbf{WAIC}}&\multicolumn{2}{c}{\textbf{p.w}}\\
&\textbf{Model (\ref{weight1})}&\textbf{Model (\ref{weight2})}&\textbf{Model (\ref{weight1})}&\textbf{Model (\ref{weight2})}\\\hline
4&92,881.1&74,730.5&5,653.2&2,637.8\\
8&92,839.3&74,363.3& 5,610.7&2,501.7\\
16&92,838.5&74,279.8&5,606.1&2,458.5\\\hline

\multirow{2}{*}{\textbf{m}}  &\multicolumn{2}{c}{\textbf{LMPL}}&\multicolumn{2}{c}{\textbf{Relative time}}\\
&\textbf{Model (\ref{weight1})}&\textbf{Model (\ref{weight2})}&\textbf{Model  (\ref{weight1})}&\textbf{Model (\ref{weight2})}\\\hline
4 & -41,340.4&-34,179.7 &1 &1.17\\
8 &-40,753.9& -34,126.2&1.65  &1.82\\
16 &-40,752.3&-34,132.1 &2.89 &3.09\\\hline
\end{tabular}
\caption{Summary of the fit of the models for $m=4,8,16$ as measured by the Watanabe-Akaike Information Criteria (WAIC) and the Log Marginal Predictive Likelihood (LMPL) statistics, as well as the relative computational time required to fit each model (compared to Model (\ref{weight1}) with $m=4$).} \label{modelfit}
\end{table}

\subsection{Covariate effects}
The estimated relative rates (RR) associated with each covariate and respiratory prescription rates are displayed in Table \ref{RR}, which presents results for both models (\ref{weight1}) and (\ref{weight2}) with $m=8$.  The value of $m=8$ was chosen as it fitted the data nearly as well as $m=16$ (the LMPL for model (\ref{weight2}) was actually better for $m=8$ compared to $m=16$) but had a much lower computational cost (see Table \ref{modelfit}). The inclusion of both models  (\ref{weight1}) and (\ref{weight2}) allows one to see the impact of this choice on the estimated results. These relative rates relate to a 1 standard deviation increase in the relevant covariate (except the indicator variable for December), as this represents a realistic long-term increase in each covariate. 

The table shows that increased PM$_{2.5}$ concentrations are associated with increased rates of respiratory prescriptions, with an estimated relative rate ranging between 1.016 and 1.022 and 95\% credible intervals that do not include the null risk of one for either model. Thus, as model (\ref{weight2}) fits the data best, an increase of 1.23$\mu g m^{-3}$ in PM$_{2.5}$ is associated with an increased prescription rate of 2.2\%. The estimated relative rate for the coarser particulate matter metric PM$_{10}$ is smaller at 1.016, with a 95\% credible interval from the adaptive model that just includes the null rate of one. Thus from model (\ref{weight2}) both particulate matter metrics exhibit some evidence of an association with increased rates of respiratory prescription, but it is the finer particles PM$_{2.5}$ that have the larger and more substantial impact, which may be because these smaller particles can travel further into the lungs. Poverty has a large impact on respiratory prescription rates, as a $\pounds 57,400$ decrease in average property price results in around a 16\%  ($1/0.861=1.161$) increased prescription rate.  The results also show that people who are white have a small increased rate of prescriptions compared to non-white people, with an estimated relative rate of 1.014 (a 1.4\% increase) for a 6.7\% increase in the white population. Finally, as suggested by Figure \ref{figure2} December has an increased rate of prescription, with an increased rate of around 23\%. Finally, these estimates differ slightly between the global and adaptive models, with for example the rate associated with PM$_{2.5}$ changing from a 2.2\% increase to a 1.6\% increase, while the rate for property price changes from 16\% to 19\%.

\begin{table}
\centering\begin{tabular}{lll}
\hline
\multirow{2}{*}{\textbf{Covariate}} &\multicolumn{2}{c}{\textbf{Relative rate}}\\
&\textbf{Model (\ref{weight1}) $m=8$}&\textbf{Model (\ref{weight2}) $m=8$}\\\hline
PM$_{2.5}$ (SD=1.23$\mu gm^{-3}$)&1.016 (1.003, 1.030) &1.022 (1.005, 1.040) \\
PM$_{10}$ (SD=1.84$\mu gm^{-3}$)&1.016 (1.002, 1.030)&1.016 (0.999, 1.035)\\
Property price (SD = $\pounds 57,400$)&0.839 (0.829, 0.847) &0.861 (0.850, 0.873)\\
Proportion white (SD = 0.067) &1.017 (1.005, 1.028)&1.014 (0.998, 1.031)\\
December &1.215 (1.205, 1.225)&1.227 (1.209, 1.246)\\\hline
\end{tabular}
\caption{Estimated relative rates and 95\% credible intervals for the covariates. The results relate to both models (\ref{weight1}) and (\ref{weight2}) with $m=8$. The numbers in brackets in the first row are standard deviations (SD) for the covariate, and are the size of the increase that the relative rates relates to.} \label{RR}
\end{table}

\subsection{Health board inequalities}
The inequality in the rate of respiratory prescriptions by health board is displayed in Figure 4, where the top panel displays the posterior median rate while the bottom panel  classifies the health boards into 3 groups depending on whether their 95\% credible intervals contain the null rate of one. The left panel shows the health board effects without covariate adjustment (i.e. based on a model with no other covariates), while the right panel shows the residual health board effects after covariate adjustment. The figure shows that Dumfries and Galloway, and  Greater Glasgow and Clyde have elevated rates with or without covariate adjustment, with increased rates without adjustment of 27\% and 19\% respectively  relative to the average in Scotland. In contrast Grampian and Orkney exhibit reduced rates relative to the Scottish average, with decreased rates of 30\% and 40\% respectively without covariate adjustment. The spatial inequality in health board rates is larger without covariate adjustment than in the residual rate after the adjustment, with ranges of $(0.716, 1.268)$ without adjustment and $(0.787, 1.230)$ with adjustment respectively.

\begin{figure}
\centering\caption{The top rows shows maps of the posterior median health board effects without (a) and with (b) covariate adjustment. The bottom row shows whether the health board effects are substantially elevated or reduced from the null rate of one (or neither), measured as whether the 95\% credible intervals include 1 or not. This is displayed without (c) or with (d) covariate adjustment.}
\label{figure4}
\begin{picture}(10,16)
\put(-1.8,0){\scalebox{0.25}{\includegraphics{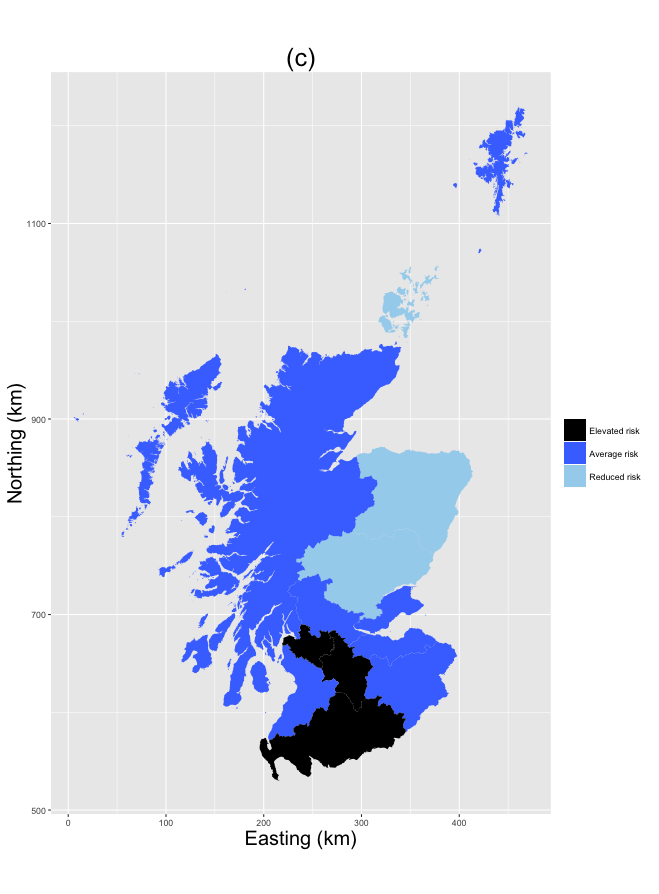}}}
\put(5.2,0){\scalebox{0.25}{\includegraphics{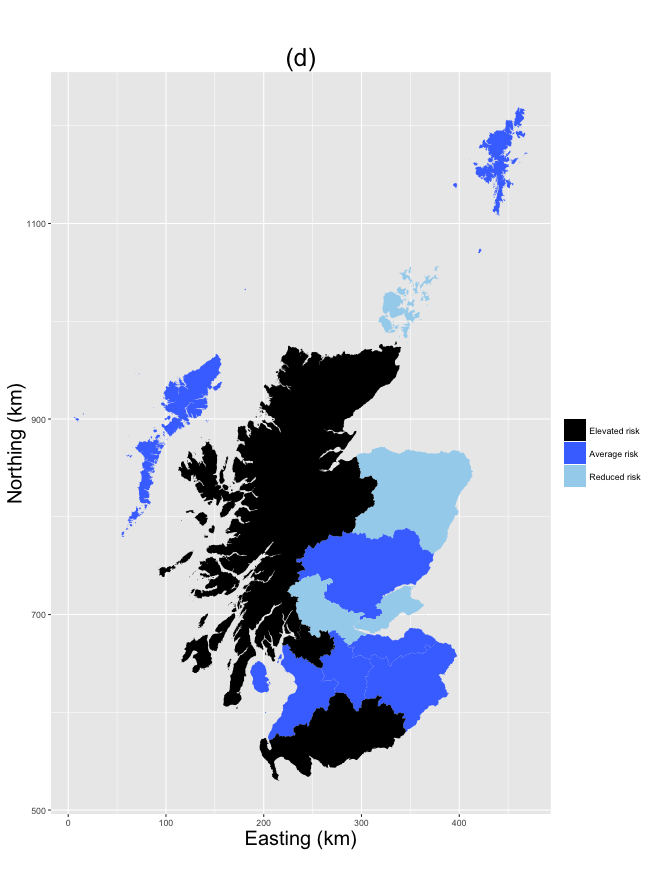}}}
\put(-2,8){\scalebox{0.25}{\includegraphics{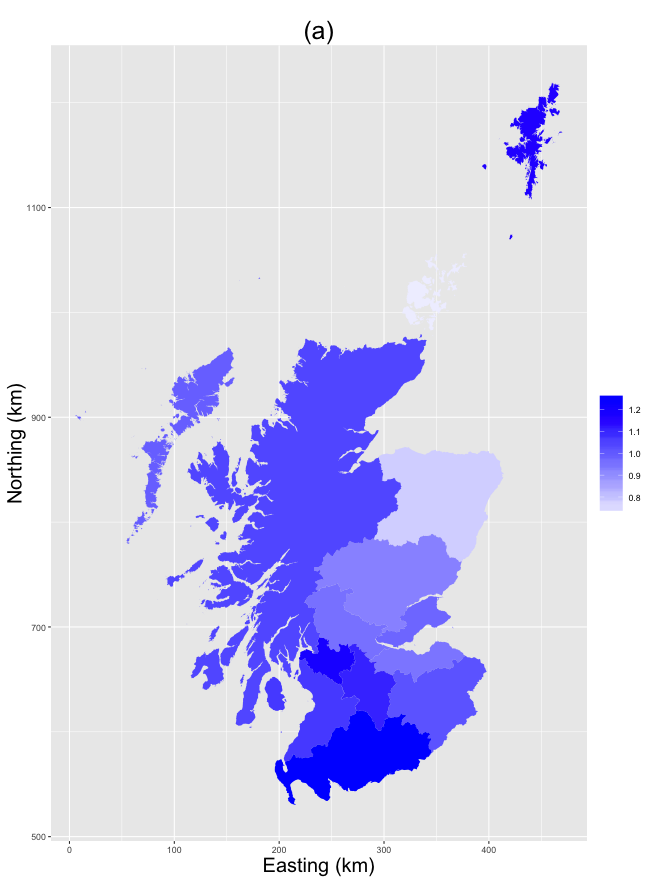}}}
\put(5,8){\scalebox{0.25}{\includegraphics{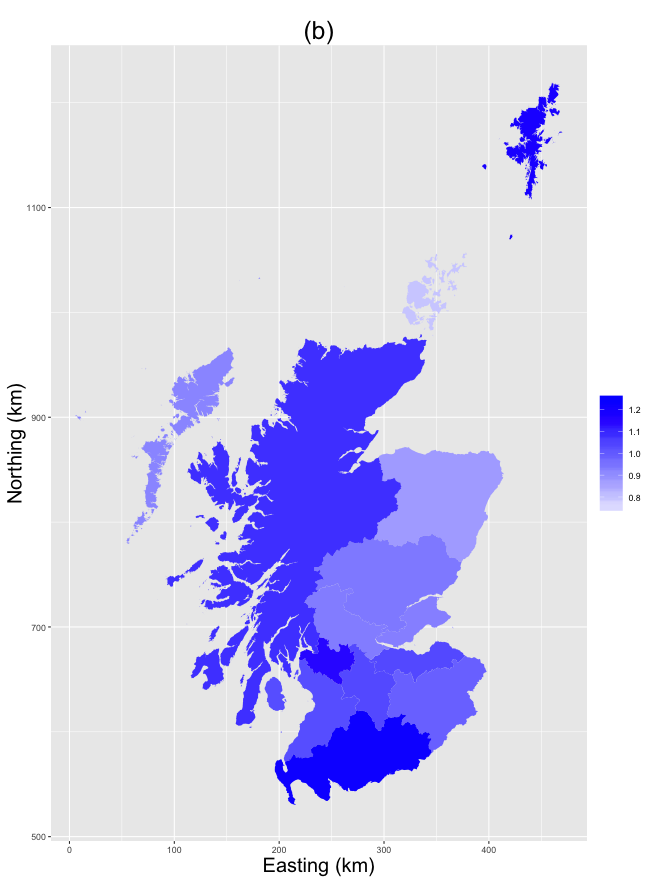}}}
\end{picture}
\end{figure}

\subsection{Localised spatial structure}
\begin{figure}
\centering\caption{Posterior median spatial autocorrelations amongst the random effects (taken from (\ref{corrandvar})) from the global smoothing model (\ref{weight1}) plotted against distance between GP surgeries. The colour shading shows the corresponding autocorrelation estimates from the adaptive smoothing model (\ref{weight2}).}
\label{figure5}
\scalebox{0.35}{\includegraphics{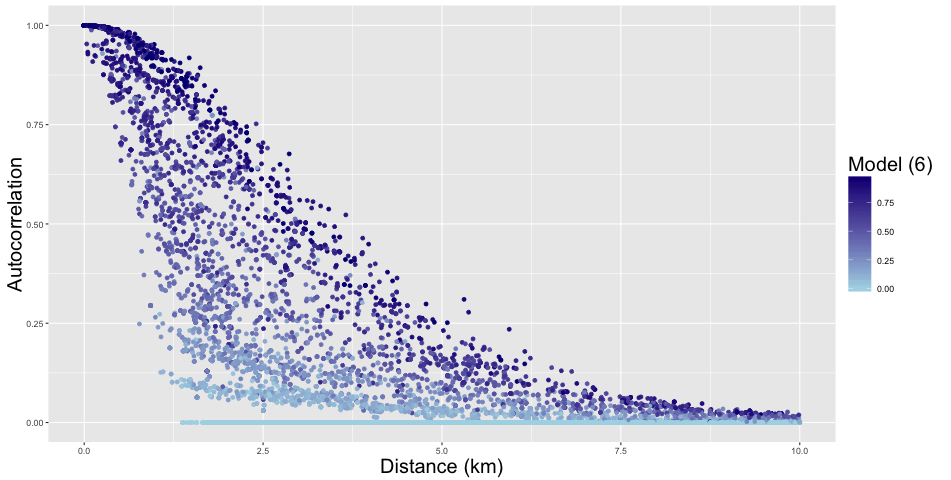}}
\end{figure}

The spatial autocorrelation structure between the random effects $(\phi_{t}(\mathbf{s}_k),\phi_{t}(\mathbf{s}_i))$ from (\ref{corrandvar}) imposed by both the global (\ref{weight1}) and adaptive (\ref{weight2}) models  is displayed in Figure \ref{figure5}, which in both cases corresponds to $m=8$. The figure displays the posterior median pairwise autocorrelation from the global model (\ref{weight1}) against distance between GP surgeries, while the colour of the points identifies the posterior median pairwise autocorrelation from the adaptive model (\ref{weight2}). The figure is truncated at 10km at which point the autocorrelation from the global model is almost zero. As expected, the autocorrelation from the global model smoothly decays as distance between the GP surgeries increases, with the range of autocorrelations at a single distance being caused by the spatial configuration of the GP surgeries (e.g. some surgeries have lots of nearby surgeries while others do not). In contrast, the colour shading shows that the adaptive model produces a much less rigid spatial distance-decay relationship, as some pairs of nearby surgeries have very low autocorrelations while other pairs further apart have much higher autocorrelations. For example, some of the points at the 10km cut-off show substantial autocorrelations around 0.75 under the adaptive model.

\section{Discussion}
This paper has developed novel Bayesian spatio-temporal  methodology for a newly released epidemiological data set on respiratory prescription counts at the GP surgery level between 2015 and 2016. Respiratory prescriptions make up 13\% of the costs of all prescriptions in Scotland, giving the analysis of the drivers of, and the spatial inequality  in, its rates high public health impact. The data are at a point rather than an areal unit level, but are non-traditional in that the locations are fixed and predictions cannot be made at unmeasured locations. A finite dimensional process-convolution approach has been proposed to model these data,  which unlike geostatistical models does not assume there is an underlying spatial process that is measurable everywhere. Our methodology extends existing process-convolution models  to account for adaptive or localised spatial structure via a tapered random weighting scheme, and freely available software is provided with the data to make this research reproducible. This adaptive model has been shown to fit the GP respiratory prescriptions data much better than a simpler global smoothing alternative, and  the analysis has shown two key findings. 

The first is the relationship between respiratory prescription rates and air pollution, as almost all of the existing literature has focused on more severe health events such as hospitalisations or deaths. Exceptions are the work by \cite{sofianopoulou2013} and \cite{blangiardo2016}, with both finding positive associations with PM$_{10}$ and NO$_{2}$ respectively. However, both studies aggregated their prescription data to non-overlapping areal units, and this study is the first to our knowledge to relate air pollution to point level GP prescription data. Our results suggest that particulate air pollution, and in particular PM$_{2.5}$, has a positive association with respiratory prescription rates, with an increase of 1.23$\mu gm^{-3}$ being associated with a 2.2\% increased rate. What is especially interesting is that these positive effects are consistent with the positive effects found for more severe respiratory outcomes such as hospitalisation or death using a range of study designs (e.g. \citealp{lee2008} and \citealp{rushworth2015}), and provides further corroborating evidence that air pollution has a substantial impact on respiratory disease and its associated health care costs. This positive effect occurs despite the fact that Scotland has relatively low particulate matter concentrations, with \cite{ricardo2016} reporting that in 2015 none of the 76 sites that monitored PM$_{10}$  in Scotland exceeded the UK Air Quality Strategy  limit of an annual mean of 40$\mu gm^{-3}$. Thus this paper adds to the evidence pool  that low levels of air pollution below legal limits are associated with adverse respiratory health.

The second key finding of this analysis is about the nature and extent of health inequalities in respiratory prescription rates between the Scottish health boards. These inequalities are largely consistent with or without covariate adjustment, the latter being the residual health board effect on respiratory prescription rates after removing the effects of known covariates. The results show the highest rates of respiratory prescription are in Dumfries and Galloway, and  Greater Glasgow and Clyde, while the lowest rates are observed in Grampian and Orkney.  The magnitude of these inequalities is substantial, as for example Dumfries and Galloway exhibit a 27\% increased rate compared to the Scottish average, while Orkney shows a 40\% decreased rate. Results quantifying these inequalities are important for two reasons. Firstly because they provide evidence of the substantial current inequality, which feeds the on-going political discussion in this area resulting from the Marmot review in England in 2010 and the recent report by \cite{nhs2016}. Secondly, they provide evidence in disparities in health care costs, which can be used to feed into the allocation to health boards of resources via the Resource Allocation Formula (\url{http://www.isdscotland.org/Health-Topics/Finance/Resource-Allocation-Formula/}). 

The very recent free availability of GP prescription data means that there is substantial scope for future methodological development and applied statistical work in this area.  The work presented here has quantified health inequalities with respect to respiratory prescribing only, and it is of clear interest to understand the inter-relationships in health board inequalities between prescribing rates for different diseases. This leads naturally to the development of a multivariate extension to the spatio-temporal process-convolution model proposed here, which could incorporate the adaptive smoothing borrowing strength over the different prescription types. From a public health agency  perspective data visualisation is key to  understanding  the inequalities and presenting a clear message to policy makers, so the development of a software package to 
allow users to interact with the inequality data and models will be a fruitful research area to pursue.

\bibliographystyle{chicago}
\bibliography{ma}
\end{document}